\documentstyle[12pt,epsf]{article}
\newcommand{\vac}{|0\rangle}
\newcommand{\low}{|G\rangle}
\newcommand{\lowp}{|G'\rangle}

\newcommand{\II}{{\cal I}}

\newcommand{\wt}{\widetilde}

\newcommand{\be}{\begin{equation}}
\newcommand{\ee}{\end{equation}}
\newcommand{\ben}{\begin{eqnarray}\displaystyle}
\newcommand{\een}{\end{eqnarray}}
\newcommand{\refb}[1]{(\ref{#1})}
\newcommand{\p}{\partial}
\newcommand{\sectiono}[1]{\section{#1}\setcounter{equation}{0}}

\begin{document}

{}~ \hfill\vbox{\hbox{hep-th/9808141}\hbox{MRI-PHY/P980858}
}\break

\vskip 3.5cm

\centerline{\large \bf SO(32) Spinors of Type I and Other Solitons}
\medskip
\centerline{\large \bf  on Brane-Antibrane Pair}

\vspace*{6.0ex}

\centerline{\large \rm Ashoke Sen
\footnote{E-mail: sen@mri.ernet.in}}

\vspace*{1.5ex}

\centerline{\large \it Mehta Research Institute of Mathematics}
 \centerline{\large \it and Mathematical Physics}

\centerline{\large \it  Chhatnag Road, Jhoosi,
Allahabad 211019, INDIA}

\vspace*{4.5ex}

\centerline {\bf Abstract}

We construct the SO(32) spinor state in weakly coupled type I string
theory as a kink solution of the tachyon field on the D-string $-$
anti-D-string pair and calculate its mass. We also give a description of
this system in terms of an exact boundary conformal field theory and show
that in this description this state can be regarded as a
non-supersymmetric D0-brane in type I string theory.  This construction
can be generalised to represent the D0-brane in type IIA string theory as
a vortex solution of the tachyon field on the membrane anti-membrane pair,
and the D-string of type I string theory as a topological soliton of the
tachyon field on the D5-brane anti- D5-brane pair.

\vfill \eject

\tableofcontents

\baselineskip=18pt

\sectiono{Introduction and Summary} \label{s1}

It has been conjectured that SO(32) heterotic string theory is dual to
type I string theory\cite{WITTEND,DABH,HULL,POLWIT}. Many tests of this
conjectured duality have been performed, including comparison of the BPS
spectrum in compactified string theories\cite{TEST}. However, even in ten
dimensions, the SO(32) heterotic string theory contains, in its
perturbative spectrum, states which are stable but are not BPS. These are
the states which transform as spinors of SO(32). The lightest states
belonging to the spinor representation of SO(32) must be stable since they
cannot decay into anything else. On the other hand they are not BPS
states, since the theory in ten dimensions has only $N=1$ supersymmetry
and hence no central charge.  Like BPS states, these stable non-BPS states
must exist at all values of the coupling, although their mass is not
determined by any known formula. Thus if the duality between the type
I and SO(32) string theory is correct then we must find these states in
the weakly coupled type I string theory as well.
This is the problem we address in this paper.

A description of this
state was proposed earlier in ref.\cite{TACH} as a
tachyonic kink solution on the D-string anti- D-string pair.
In section \ref{s2} of the paper we discuss this construction in
detail. It is well known that on the world volume of the D-string
anti- D-string pair there is a tachyonic mode coming from open string
with one end on the D-string and the other end on the anti-
D-string\cite{GREEN,BANSUS,GRGUP,LIF}.
At the minimum of the tachyon potential, the negative
contribution to the energy density from the tachyon potential exactly
cancels the positive contribution from the tension of the D-string
and the anti- D-string\cite{TACH}. Since
in type I theory the tachyon is a real field, and
furthermore, the dynamics
is invariant under a change of sign of the tachyon field, the
minimum of the potential comes in pairs $\pm T_0$. Thus we can construct
a tachyonic kink solution on the D-string anti- D-string pair
interpolating between the degenerate vacua at $\pm T_0$. We show
that this state carries SO(32) spinor quantum numbers. Furthermore its mass
is finite since far away from the core of the soliton the energy density
vanishes.  Thus this is the ideal candidate for the state
we are looking for.

The mass of this state can be easily shown to be proportional to the
inverse coupling ($g^{-1}$) of the type I theory by a simple scaling
argument. However computing the constant of proportionality directly
requires the detailed knowledge of the tachyon potential which is
not available to us. We use an indirect method for computing this
coefficient by considering the case where the type I theory has been
compactified on a circle of radius $R$, and the D-string anti- D-string
pair is wrapped along that circle. We show that in order
to get an SO(32) spinor state we must impose
an anti-periodic boundary condition on the tachyon, and as a result
the zero momentum mode of the tachyon is absent.
As we reduce the radius of the
circle, the contribution to the effective mass$^2$ of various modes
from the momentum
along the circle increases, and one finds that at a critical
radius $R=(1/\sqrt 2)$ the
tachyonic mode ceases to be tachyonic. In fact at this point
the lowest momentum mode of the tachyon represents an
exactly marginal deformation. We interprete this critical radius as the
point where the SO(32) spinor state becomes degenerate with the
trivial vacuum of the D-string anti- D-string pair where the tachyon
vanishes. Since the mass of the wrapped D-string anti- D-string pair
in the absence of tachyon vacuum expectation value (vev) can be
easily computed in the weak coupling limit, this gives an indirect
measurement of the mass of the spinor state. Although this gives us
the mass only at the critical radius, we give a
heuristic argument showing that the mass is independent of
the radius and hence also represents the mass of the spinor state
at infinite radius.

This `target space viewpoint' gives an intuitive description
of the SO(32) spinor state; however it does not give us an exact conformal
field theory description of the system. This is the problem we
address in sections \ref{s3} and \ref{s4}. In section \ref{s3} we work
at the critical radius $R=(1/\sqrt 2)$, and study the effect of switching
on the marginal deformation associated with `tachyon' condensation.
Parametrizing the tachyon vev at this critical radius by an appropriate
parameter $\alpha$ we compute the spectrum of open string states on
the D-string anti- D-string system at arbitrary $\alpha$. In particular
we find that $\alpha$ is a periodic variable with periodicity $2$.
At any value of $\alpha$ other than zero (mod 2)
the mode representing the
freedom of separating the string anti-string pair becomes massive,
showing that the pair is bound. In fact the mass of this mode is
maximum at $\alpha=1$ showing that at this point the pair is
maximally bound. Also for all $\alpha\ne 0$ there is a zero mode
representing the freedom of translating the solution along the
compact direction, indicating the fact that the translation invariance
of the original string anti-string configuration along the compact
direction is broken as we
switch on the tachyon vev.

At $R=(1/\sqrt 2)$ all values of $\alpha$ represent configurations
of same
mass, and hence we cannot unambiguously identify the point which
corresponds to the spinor state, {\it i.e.} which will correspond to
the minimum of the potential when we increase $R$ beyond this critical
radius. This is the problem we address in section \ref{s4} by studying
the effect of switching on the perturbation in the closed string
sector corresponding to the radius deformation. We find that the
`tachyonic mode' develops a one point function at all $\alpha$
except at $\alpha=0$ and 1. Thus only these two points correspond
to the extremum of the tachyon potential. Furthermore, the `tachyonic
mode' becomes tachyonic at $\alpha=0$ as expected, but acquires
positive mass$^2$ at $\alpha=1$. This shows that the point $\alpha=1$
represents a stable minimum of the potential, and
hence represents the SO(32)
spinor state that we have been looking for.

One can now examine the limit $R\to \infty$. This is best
studied using a different set of world-sheet bosonic and fermionic
fields, related to the original bosonic and fermionic fields
representing cooordinate along the compact direction
by a series of bosonization,
fermionization and duality transformation. The net result is that
in terms of the new bosonic coordinate, the soliton at $\alpha=1$
can be represented as a D0-brane, with Dirichlet boundary condition
in all directions, and the radius of the new boson goes to $\infty$
as $R\to \infty$. This gives an exact world-sheet description
of the SO(32) spinor state of type I string theory and allows us to
calculate its mass. This is same as the mass calculated in
section \ref{s2}. The $R$ independence of the mass. which was a
crucial assumption in the analysis of section \ref{s2}, can be
easily understood by noting that the mass of a D0-brane situated
on a compact circle does not depend on the radius of the circle.

In section \ref{s5} we consider other solitons on brane anti-brane
system. The first example we study is a membrane anti-membrane pair
of type IIA string theory. There is a complex tachyon field living on
this system, and hence we expect its vacuum manifold (minimum of
the potential) to be a circle $S^1$. Since $\pi_1(S^1)$ is non-trivial,
we can now construct topologically stable tachyonic vortex solution
on the world-volume of the membrane anti-membrane pair. By analysing
the quantum numbers carried by this soliton we can identify it as
the D0-brane of type IIA string theory. This construction can be
easily generalized to represent a D$p$-brane of type II string theory
as a vortex solution on the D-$(p+2)$-brane anti- D-$(p+2)$-brane pair.

The other example that we study in this section is that of a D5-brane
anti- D5-brane pair in type I string theory. The tachyon field
living on this system is represented by a $2\times 2$ matrix, and
the vacuum manifold for this tachyon field can be shown to be $S^3$.
Thus we can construct a topologically stable string like solution
on this $(5+1)$ dimensional world volume with the property that
the asymptotic boundary of the string, which in this case is $S^3$,
wraps around the vacuum manifold with unit winding number. A detailed
analysis of this solution shows that it carries the same quantum numbers
as those of
a D-string of type I string theory. Thus it gives a different
representation of the D-string in this theory.

Our analysis also throws light on a related problem discussed in
refs.\cite{BOUND,BERGAB}. If we consider an orbifold plane of type
IIB string theory of the form $R^{5,1}\times
(R^4/(-1)^{F_L}\cdot\II_4)$, where $\II_4$ denotes
the $Z_2$ transformation
that reverses the sign of all coordinates along $R^4$ and
$(-1)^{F_L}$ denotes the reversal of sign of all Ramond sector
states on the left, then the twisted sector states contain a massless
U(1) gauge field living on the orbifold plane $R^{(5,1)}$ located
at the origin of $R^4$. It is a prediction of $S$-duality that there
must be stable non-BPS states living on the orbifold plane which
are charged under this U(1) gauge field\cite{NONBPS}.
In ref.\cite{BOUND} this
state was identified as a half soliton living on a D-string
anti- D-string pair transverse to the orbifold plane, obtained by
modding out the kink solution described in section \ref{s2} (regarded as
a configuration in type IIB string theory rather then in type I
theory) by a reflection around its origin. On the other hand,
ref.\cite{BERGAB} attempted to describe the same state as a
non-supersymmetric D0-brane, and constructed the boundary state
describing this system. Our present analysis shows that these are
different descriptions of the same physical object.\footnote{The
apparent discrepancy in the mass of the state calculated in the
two approaches\cite{BERGAB} seems to be due to the
different units that have been used in the two papers.}

We conclude this section by mentioning some related developments
which have taken place during the recent years. Various examples of
tachyon condensation in string theory have
been studied in \cite{SRED}. Stable non-BPS states in the
context of supersymmetric field theory have been analyzed in
\cite{STRASS,NONFI,BERKOL}. In particular, \cite{STRASS}
discusses non-BPS states carrying $Z_2$ quantum numbers
similar to the states considered here. Other aspects of tachyons
in non-supersymmetric string theories have been discussed in
\cite{NONSUP}.

\sectiono{SO(32) Spinor State as a Tachyonic Kink on the D-string Anti-
D-string Pair} \label{s2}

Let us consider type I string theory compactified on a circle $S^1$ of
radius $R$ and a D-string of type I wrapped on $S^1$. It is known that the
world volume theory on the D-string is identical to that of the
fundamental heterotic string in SO(32) heterotic string
theory\cite{POLWIT}. In particular the U(1) gauge field on the D-string is
projected out by the requirement of world-sheet parity invariance.  There
is however a $Z_2$ subgroup of this U(1) which survives this projection,
and hence we can introduce a $Z_2$ Wilson line on the $S^1$. Since the end
of a fundamental string ending on the D-string is charged under the $Z_2$,
in the presence of a Wilson line the wave function of an open string with
one end on the D-string will pick up an extra $-$ sign under a $2\pi R$
translation along $S^1$.

The open string states with one end on the D-string and the other end on
any of the 32 9-branes filling up the space-time provide the 32 fermionic
degrees of freedom on the D-string world-volume. Due to the possibility of
putting the $Z_2$ Wilson line on the D-string, these fermions can satisfy
either periodic or anti-periodic bounday condition along $S^1$. If we put
anti-periodic boundary condition, then there are no fermion zero modes,
and the states are in the scalar conjugacy class of SO(32). On the other
hand if we put periodic boundary condition on these fermions, then there
are zero modes of the 32 fermions, and the quantization of these zero
modes give a ground state (and hence all excited states) in the spinor
conjugacy class of SO(32).

Thus if we consider a system of a D-string $-$ anti- D-string\footnote{An
anti- D-string is simply a D-string with opposite orientation.} pair of
type I, both wrapped on $S^1$, and put $Z_2$ Wilson line on one of them,
the state will belong to the spinor conjugacy class of SO(32), and will
not carry any other conserved charge as the Ramond-Ramond (RR) two form
charges of the D-string and the anti- D-string will cancel each other.
Thus this state has the correct quantum numbers, and we expect the lowest
mass state in this sector to represent the lightest SO(32) spinor state of
type I string theory. Naively, the mass of this state is given by, \be
\label{e1} 2\cdot 2\pi R \cdot T_D\, , \ee where $T_D$ represents the
tension of a single D-string:
\be \label{e2} T_D = {1\over 2\pi g} \, ,
\ee
$g$ being the string coupling constant. (We are working in units where we
have set $\alpha'=1$. Also we shall view type I string theory as the
result of modding out type IIB theory by the world-sheet parity
transformation $\Omega$, and use type IIB units before projection
under $\Omega$ to calculate all masses.)
{}From eq.\refb{e1} we see that as $R\to\infty$, the
mass of the D-string anti-D-string pair goes to infinity. Thus although
this configuration has the right quantum numbers, it cannot represent the
state we are looking for.

The situation improves upon noting that on the D-string anti- D-string
world volume there is a tachyon field\cite{GREEN,BANSUS,GRGUP,LIF},
representing the fact that we are
actually sitting at the top of a potential well. Thus in order to find the
lowest energy configuration with these quantum numbers we must allow the
tachyon field to acquire a vacuum expectation value and go to the minimum
of its potential. Let $T$ denote the tachyon field living on the D-string
anti- D-string world-volume, $g^{-1}V(T)$ denote the tachyon
potential,\footnote{We normalise various fields so that the world-volume
action on the D-string has an overall factor of $g^{-1}$ in front, and
does not have any other $g$ dependence.} and $T_0$ be the point where
$V(T)$ is at its minimum. It was argued in \cite{TACH} that, \be
\label{e3} g^{-1} V(T_0) + 2 T_D =0 \, , \ee so that at the minimum of the
tachyon potential the net energy per unit length of the D-string anti-
D-string pair vanishes. Note that since the tachyon arises in the open
string sector with one end on the D-string and the other end on the anti-
D-string, it is odd under the $Z_2$ gauge transformation associated with
either of the strings, and hence $V(T)$ is symmetric under $T\to -T$. Thus
the locations of the minimum of $V(T)$ come in pairs $\pm T_0$.

Naively one would think that the ground state of the D-string anti-
D-string system that we have been considering will correspond to a
constant tachyon configuration $T(x)=T_0$, where $x$ denotes the
coordinate along $S^1$. However since we have put a $Z_2$ Wilson line on
one of the D-strings, and since the tachyon is odd under this $Z_2$ gauge
transformation, it must be anti-periodic under translation by $2\pi R$
along $S^1$. This shows that $T=T_0$ is not an allowed configuration.
Instead, for large $R$, the minimum energy configuration will correspond
to a kink solution such that\footnote{Here we have taken the core of the
soliton to be located at $x=0$.} \be \label{e4} T(x) \to -T_0 \quad
\hbox{for} \quad x<<0,\qquad T(x) \to T_0 \quad \hbox{for} \quad x>>0, \,
, \ee so that the tachyon satisfies the anti-periodic boundary condition:
\be \label{e5} T(\pi R) = - T(-\pi R) \, .  \ee This configuration has
finite mass in the $R\to \infty$ limit as a consequence of \refb{e3} since
the energy density vanishes far away from the core. In fact, far away from
the core the D-string anti- D-string configuration is indistinguishible
from the vacuum since it does not carry any net charge or energy density,
so the configuration is localised near $x=0$ and represents
{\it a particle
like state} in the ten dimensional type I string theory. The total mass of
this state can be computed by integrating the energy density along $x$.
Since the dependence on the string coupling $g$ of this energy density is
through an overall multiplicative factor of $g^{-1}$, we can immediately
conclude that the mass of this soliton is given by \be \label{e6}
m_{spinor} = C / g\, , \ee where $C$ is a numerical constant. Explicit
computation of $C$ requires a detailed knowledge of the tachyon potential,
but we shall compute it through an indirect argument later.

Thus we see that {\it the SO(32) spinor state in type I string theory can
be identified as the tachyonic kink on the D-string anti- D-string pair.}
Although our argument based on circle compactification already shows that
this state transforms in the spinor representation of SO(32), we can also
see it directly without the circle compactification.\footnote{The
following investigation was suggested by N. Seiberg.} On the world-volume
of the D-string there are 32 left-moving Majorana fermions $\chi^i_{(1)}$,
and on the world volume of the anti- D-string there are 32 right-moving
Majorana fermions $\chi^i_{(2)}$ ($1\le i\le 32$). Since under the
$Z_2\times Z_2'$ gauge symmetry on the world volume of the D-string anti-
D-string system the fields $\chi^i_{(1)}$, $\chi^i_{(2)}$ and $T$ carry
charges $(-1,1)$, $(1,-1)$ and $(-1,-1)$ respectively, the coupling
between these fields must have the form:
\be \label{e7} L_{int} = i f(T)
\chi^i_{(1)} \chi_{(2)}^i + \cdots \, ,
\ee
where $f(T)$ is an odd
function of $T$. $\cdots$ in the above equation denote terms quartic and
higher order in the fermion fields. Thus the relevant
quadratic part of the
fermionic lagrangian is given by:
\be \label{e8} L_{tot} = i
[\chi^i_{(1)} (\p_t -\p_x) \chi^i_{(1)} + \chi^i_{(2)} (\p_t +\p_x)
\chi^i_{(2)} ] + i f(T) \chi^i_{(1)} \chi^i_{(2)}\, .
\ee
In general the
kinetic terms could be renormalized by even functions of $T$, but these
could be absorbed into a redefinition of the fields $\chi^i_{(1)}$ and
$\chi^i_{(2)}$. We shall now look for fermion zero modes by looking at the
time independent equations of motion derived from the lagrangian
\refb{e8}:
\be \label{e9} \p_x \pmatrix{\chi^i_{(1)}\cr \chi^i_{(2)}} =
f(T) \sigma_1 \pmatrix{\chi^i_{(1)}\cr \chi^i_{(2)}} \, ,
\ee
where
$\sigma_i$ denote the Pauli matrices:
\be \label{e10} \sigma_1 =
\pmatrix{0 & 1\cr 1 & 0}\, , \qquad \sigma_2 = \pmatrix{0 & -i\cr i & 0}\,
, \qquad \sigma_3=\pmatrix{1 & 0\cr 0 & -1} \, .
\ee
Let us now assume
that $f(T_0)>0$. Since $f(T)$ is an odd function of $T$, this would imply
that $f(-T_0)<0$. Using this we get the following normalizable solutions
of eq.\refb{e9}
\be \label{e11}
\pmatrix{\chi^i_{(1)}\cr \chi^i_{(2)}} =
\chi_0^i \exp(\sigma_1 \int_0^x f(T(x')) dx')\pmatrix{1 \cr -1}\, .
\ee
$\chi_0^i$ are independent constants. If $f(T_0)<0$, then the vector
$\pmatrix{1\cr -1}$ in \refb{e11} is replaced by $\pmatrix{1\cr 1}$.
In either case we get 32 fermionic zero modes $\chi_0^i$
transforming in the vector representation of SO(32). Quantization of these
zero modes gives states in the spinor as well as in the conjugate spinor
representation of SO(32). However, since the zero modes are odd
under the combined action of $Z_2$ and $Z_2'$ generators which leave $T$
and hence the background solution invariant, by demanding that the states
are invariant under this diagonal $Z_2$ group leaves us only with states
in the spinor representation of the SO(32) group. This shows that the
tachyonic kink solution indeed has the right quantum numbers.

Let us now turn to the indirect argument determining the value of the
numerical coefficient $C$ appearing in the mass formula \refb{e6}. For this
we start with type I theory on a circle $S^1$ of radius $R$, and
consider the original configuration, namely a D-string anti- D-string
pair wrapped on $S^1$ with a $Z_2$ Wilson line on one of the strings.
Since the tachyon field living on the world volume of this system must
satisfy anti-periodic boundary condition along $S^1$, it admits a mode
expansion of the form:
\be \label{e12}
T(x,t) = \sum_{n\in Z} T_{n+{1\over 2}}(t) e^{i {n+{1\over 2}\over R} x}\, .
\ee
Note that the zero momentum mode is absent. The effective mass$^2$ of
$T_{\pm(n+{1\over 2})}$ is given by
\be \label{e13}
m_n^2 = {(n+{1\over 2})^2\over R^2} -{1\over 2}\, ,
\ee
where $-1/2$ is the mass$^2$ of the tachyon field on the
infinite D-string
anti- D-string pair. From eq.\refb{e13} we see that $m_n^2\ge 0$ for
all $n$ if
\be \label{e14}
R \le {1\over \sqrt 2}\, .
\ee
In particular this shows that at $R={1\over \sqrt 2}$ the
modes $T_{\pm{1\over 2}}$ represent
marginal deformation. In fact as will be shown in the next section,
this mode actually represents an exactly marginal deformation.
Thus we arrive at the following intuitive picture.
\begin{itemize}
\item For $R>(1/\sqrt 2)$ the total mass of the D-string anti- D-string pair
before tachyon condensation is larger than the mass $m_{spinor}$ of the spinor
state, signalling the presence of a tachyonic mode
in the D-string anti- D-string system:
\be \label{e24}
4 \pi R T_D > m_{spinor} \qquad \hbox{for} \qquad R > {1\over \sqrt 2}\, .
\ee
\item For $R<(1/\sqrt 2)$ the total mass of the D-string anti- D-string pair
before tachyon condensation is smaller than the
mass $m_{spinor}$ of the spinor
state, signalling the absence of
a tachyonic mode in the D-string anti- D-string
system:
\be \label{e25}
4 \pi R T_D < m_{spinor} \qquad \hbox{for} \qquad R < {1\over \sqrt 2}\, .
\ee
This also shows that in this range of $R$ the spinor state becomes
unstable against decay into a D-string anti- D-string pair.
\item For $R=(1/\sqrt 2)$ the total mass of the D-string anti- D-string pair
before tachyon condensation is
equal to the mass $m_{spinor}$ of the spinor
state, signalling the presence of a marginal deformation
connecting the two states.
\be \label{e26}
4 \pi R T_D = m_{spinor} \qquad \hbox{for} \qquad R = {1\over \sqrt 2}\, .
\ee
\end{itemize}
This gives
\be \label{e27}
m_{spinor} = \sqrt 2/g\, .
\ee
Comparing with eq.\refb{e6} we get
\be \label{e28}
C=\sqrt 2\, .
\ee

The above analysis determines the value of
$m_{spinor}$ at $R=(1/\sqrt 2)$. Since
we are interested in determining $m_{spinor}$ for $R=\infty$,
we have implicitly
assumed that $m_{spinor}$ does not depend on $R$, {\it i.e.} it does not get
affected by compactification of the theory on a circle. We shall now give an
indirect argument in favour of this, and leave a more detailed analysis of this
question to section \ref{s4}. In order to study the effect of compactification
on the mass of the spinor state we can estimate the interaction energy between
the spinor and its (infinite set of) images situated at distance $2\pi n R$ on
the $X$-axis for all integer $n$. In particular the gravitational interaction
in this case is of order
\be \label{e29}
g^2 \cdot g^{-1} \cdot g^{-1} \sim 1 \, ,
\ee
where the $g^2$ factor arises from the Newton's
constant and the factors of $g^{-1}$
arise from the masses of each state in the interacting pair.
Since the right hand
side of \refb{e29} is much smaller than
$g^{-1}$ $-$ the mass of the spinor state
as computed earlier $-$ we expect that the
gravitational interaction between the spinor
and its images will not significantly modify the mass of this state. Similar
argument can be given for electromagnetic interactions
as well. Thus we conclude
that the mass of the spinor state is not significantly modified due to
interaction with its images, and hence \refb{e27} represents the
mass of this state in the
infinite radius limit as well.

Finally, note that although we would have expected the
classical soliton representing the SO(32) spinor state in (9+1)
dimension to be invariant under the SO(9) rotation group, the
soliton that we have constructed only has manifest SO(8)
symmetry since the direction along which the original D-string
anti- D-string pair lies is somewhat special. However, we shall see in
section \ref{s4} that the boundary conformal field theory describing this
state does have the full SO(9) rotational invariance.

\sectiono{Conformal Field Theory at the Critical Radius} \label{s3}

Let $X$ denote the scalar field
corresponding to coordinate $x$ along the $S^1$ and $\psi$, $\wt\psi$ its
right- and left-moving fermionic partners on
the world-sheet of the fundamental
string. At this particular radius the conformal field theory of the bosonic
field $X$ is equivalent to that of a pair of right-moving fermionic fields
$\xi$, $\eta$ and a pair of left-moving fermion fields $\wt\xi$,
$\wt\eta$\cite{GINSP}.
If we decompose $X$ into its left- and right-moving parts as
\be \label{e15}
X = X_L + X_R\, ,
\ee
then the bose-fermi relation takes the form:
\be \label{e16}
e^{i\sqrt 2 X_R} = {1\over \sqrt 2} (\xi + i\eta)\, , \qquad
e^{i\sqrt 2 X_L} = {1\over \sqrt 2} (\wt\xi + i\wt\eta)\, .
\ee
We can find another representation of the same conformal field theory by
rebosonising the fermions as follows:
\be \label{e22}
{1\over \sqrt 2} (\xi + i\psi) = e^{{i \sqrt 2} \phi_R}\, , \qquad
{1\over \sqrt 2} (\wt\xi + i\wt\psi) = e^{{i \sqrt 2} \phi_L}\, .
\ee
$\phi$ represents a free bosonic field with radius $1/\sqrt 2$.  There is
a third representation in which we use a slightly different rebosonization:
\be \label{e22a}
{1\over \sqrt 2} (\eta + i\psi) = e^{{i \sqrt 2} \phi'_R}\, , \qquad
{1\over \sqrt 2} (\wt\eta + i\wt\psi) = e^{{i \sqrt 2} \phi'_L}\, ,
\ee
where $\phi'$ is another scalar field of radius $1/\sqrt 2$. For later
use we list here the operator product expansions, and the relations
between the currents of free fermions and bosons:
\be \label{eyy1}
\psi(z)\psi(w)\simeq \xi(z)\xi(w)\simeq\eta(z)\eta(w)\simeq
{i\over z-w} \, ,
\ee
\be \label{eyy2}
\p X_R(z) \p X_R(w) \simeq \p \phi_R(z) \p\phi_R(w) \simeq
\p\phi'_R(z)\p\phi'_R(w) \simeq -{1\over 2(z-w)^2}\, ,
\ee
\be \label{eyy3}
\psi\xi = i\sqrt 2 \p \phi_R\, , \qquad \eta\xi = i\sqrt 2 \p X_R\, ,
\qquad \psi\eta = i\sqrt 2 \p \phi'_R\, .
\ee
Here $\simeq$ denotes equality up to non-singular terms. There are also
similar relations involving the left-moving currents.

Since $X$ satisfies Neumann boundary condition $X_L=X_R$
at the boundary of the world
sheet, this translates to the Neumann boundary condition on the
fermions:\footnote{These
boundary conditions are written in the coordinate system where the open string
world sheet is represented as the upper half plane.}
\be \label{e17}
\xi_B=\wt\xi_B\, , \qquad \eta_B=\wt\eta_B\, ,
\ee
where the subscript $B$ denotes the boundary
values of the fields. On the other
hand in the NS sector, $\psi$ satisfies the
boundary condition
\be \label{e17a}
\psi_B=\wt \psi_B\, ,
\ee
at both boundaries. Thus we see from \refb{e22a} that
$\phi$ and $\phi'$ both satisfy
Neumann boundary condition at both ends. However, in the Ramond sector, at one
of the two ends of the open string, $\psi$ satisfies:
\be \label{e17b}
\psi_B=-\wt \psi_B\, .
\ee
This translates to the fact that $\phi$ and $\phi'$ satisfy Dirichlet boundary
condition at one end and Neumann boundary condition at the other end.

Up to overall numerical factors, the
vertex operators corresponding to the states $T_{\pm{1\over 2}}$ are given
in the $-1$ picture as\cite{FMS}:
\be \label{e18}
V^{(-1)}_\pm = \mp i e^{-\Phi_B} e^{\pm(i/ \sqrt 2)
X_B} \otimes \sigma_1 \, ,
\ee
and in the zero picture as
\be \label{e19}
V^{(0)}_\pm = \psi_B e^{\pm (i/\sqrt 2)X_B} \otimes \sigma_1\, .
\ee
$\Phi$ denotes the bosonized ghost\cite{FMS}.
$\sigma_1$ is the $2\times 2$ Chan-Paton factor representing that the tachyon
originates in the open string sector with two ends of the open string on two
different strings. The overall multiplicative factor of $\mp1$ in \refb{e18}
is a matter of convention.  Let
\be \label{e20}
V_T = {1\over \sqrt 2} (V_+ + V_-)\, ,
\ee
represent vertex operators for the real component of $T_{1/2}$.\footnote{We
could have chosen to work with any particular direction in the
complex $T_{1/2}$
plane; we choose the real direction for definiteness.} Using
eqs.\refb{e16}-\refb{e20} we get
\be \label{e21}
V^{(-1)}_T = e^{-\Phi_B} \eta_B\otimes \sigma_1, \qquad V^{(0)}_T =
\psi_B\xi_B \otimes \sigma_1\, .
\ee
We can now take
$\phi$, $\eta$ and $\wt\eta$ as independent fields, and represent the tachyon
vertex operator in the zero picture as:
\be \label{e23}
V^{(0)}_T = {i\over \sqrt 2}
\p \phi_B \otimes \sigma_1\, ,
\ee
for Neumann boundary condition on $\phi$. $\partial$ denotes tangential
derivative along the boundary.
This shows that switching on the tachyon vev corresponds to switching on a
U(1) Wilson line along the bosonic direction $\phi$.
On the other hand, for Dirichlet boundary condition on $\phi$, we may
rewrite \refb{e21} as
\be \label{ekk1}
V_T^{(0)} = {i\over \sqrt 2} (\p\phi_R-\p\phi_L)_B \otimes \sigma_1
\ee
Switching on the tachyon vev now corresponds to changing the boundary
condition from $\phi=0$ to $\phi=$constant. Clearly, both \refb{e23}
and \refb{ekk1} represent marginal deformation.

Before we study the effect of switching
on such a tachyon vev on the spectrum of open strings, let us
study the effect of various projections on the open string spectrum
before switching on the tachyon vev.
There are several discrete transformations
under which the open string states are required to be invariant. Each
such transformation
includes transformation on the fields at the boundary, on the SL(2,R)
invariant vacuum in the NS sector, as well as on the Chan-Paton
factor which we shall represent by a $2\times 2$ matrix $\Lambda$.
These are as follows:
\ben \label{en1}
(-1)^F &:& X_B\to X_B, \quad \psi_B\to -\psi_B, \quad
X^\mu_B\to X^\mu_B,
\quad \psi^\mu_B\to -\psi^\mu_B, \nonumber \\
&& \vac \to -\vac,
\quad \Lambda\to \sigma_3 \Lambda
\sigma_3, \nonumber \\ \cr
h &:& X_B\to X_B+{2\pi\over\sqrt 2}, \quad \psi_B\to \psi_B, \quad
X^\mu_B\to X^\mu_B, \quad \psi^\mu_B\to \psi^\mu_B, \nonumber \\
&& \vac\to\vac, \quad \Lambda\to \sigma_3\Lambda \sigma_3,
\nonumber \\
\een
where $x^\mu$ ($1\le\mu\le 8$) denote the non-compact directions transverse
to the string.
In $(-1)^F$ the conjugation by $\sigma_3$ reflects the fact that
the open strings with two ends on two different strings, represented by
off diagonal Chan Paton
matrices, have opposite GSO projection compared to open strings
with both ends on the same string. The $-$ sign in the transformation law
of the vacuum represents the fact
that the NS sector ground state for open strings with both ends on the
same string is odd under $(-1)^F$. $h$ represents translation by
$2\pi /\sqrt 2$ along $x$, and
the conjugation of $\Lambda$ by $\sigma_3$ in the transformation law of $h$
is a reflection of the $Z_2$
Wilson line present along one of the strings which make open strings with
off-diagonal Chan-Paton factors anti-periodic along $x$ instead of
peridic. Using eqs.\refb{e16} we see that the action of these transformations
on the fields $\xi_B$ and $\eta_B$ at the boundary are given by:
\be \label{en2}
(-1)^F \quad : \quad \xi_B\to \xi_B, \quad \eta_B\to\eta_B, \qquad
h\quad : \quad
\xi_B\to -\xi_B, \quad \eta_B\to -\eta_B
\ee
and hence, using eq.\refb{e22}, \refb{e22a}
we get the following action on $\phi$, $\phi'$:
\be \label{en3}
(-1)^F \quad : \quad \phi_B\to -\phi_B, \quad \phi'_B\to -\phi'_B,
\qquad h \quad :\quad \phi_B\to
{2\pi\over \sqrt 2} - \phi_B\, ,
\quad \phi_B'\to {2\pi\over \sqrt 2} - \phi_B'\, .
\ee
Besides $(-1)^F$ and $h$,
the states are also required to be invariant under the
world-sheet parity transformation $\Omega$ which we have not listed
here.

Let us now study the effect of switching on the tachyon vev.
Since
we have seen that the Ramond sector of the open strings correspond to putting
Neumann boundary condition at one end and Dirichlet boundary condition at the
other, and hence are forced to carry zero momentum and
winding along $\phi$, we see
immediately that the tachyon vev, representing translation along
$\phi$ at the Dirichlet end, and Wilson line along $\phi$ at the
Neumann end,  does not affect
the Ramond sector states.
On the NS sector states with Neumann boundary condition at both
ends, the tachyon vev
corresponds to switching on a Wilson line along $\phi$. Let us
parametrize this Wilson line by
\be \label{en3a}
\exp(i\int_0^{2\pi/\sqrt 2} A_\phi d\phi) =
\exp (i\alpha\pi\sigma_1/2)\, ,
\ee
corresponding to
\be \label{exx1}
A_\phi ={\alpha\over 2\sqrt 2} \sigma_1\, .
\ee
$\alpha$ is a parameter labelling the strength of the tachyon vev. Typically
the effect of switching on the Wilson line is to shift the quantization rule
for the momentum along $\phi$ for states
which are charged under the corresponding
gauge field. In particular, for a state carrying $q$ units of charge under
$\sigma_1$, the $\phi$ momentum gets shifted by an amount
\be \label{exx2}
{\alpha\over 2\sqrt 2} q\, .
\ee
Since
the $2\times 2$ identity matrix $I$
and the matrix $\sigma_1$ commute with the Chan-Paton factor $\sigma_1$
associated with the tachyon vertex operator, we see that states with
Chan Paton factors proportional to $I$ or $\sigma_1$ are
neutral under the Wilson line and hence their masses are not affected
by the tachyon vev. This leaves us with the NS sector states  with
Chan Paton factors $\sigma_3$ and $\sigma_2$ respectively. To study the effect
of the tachyon vev on states from these sectors, let us note that the
Chan Paton factors:
\be \label{en4}
\sigma_3\mp i\sigma_2
\ee
carry charges $\pm2$ respectively under the generator $\sigma_1$.
In the presence of the tachyon vev \refb{en3a} the $\phi$ momentum
quantization rule of these
states get shifted by an amount:
\be \label{en5}
\pm {\alpha\over \sqrt 2}\, .
\ee
Since $\phi$ has radius $(1/\sqrt 2)$, this shows that $\alpha$ is a
periodic variable with period 2.
It will be more convenient for our analysis to use the
fermionic language where we use the fermionic fields $\psi_B$, $\xi_B$ and
$\eta_B$ as independent boundary fields. Defining
\be \label{en6}
\chi_B = {1\over \sqrt 2} (\xi_B + i\psi_B) = e^{i\phi_B/\sqrt 2}\, ,
\ee
we can represent the effect of the tachyon vev in the sectors
$\sigma_3\pm i\sigma_2$ by taking the mode expansions of $\chi_B$,
$\chi_B^\dagger$ as follows:
\ben \label{en7}
\chi_B &=& \sum_{n\in Z} \chi_{n+{1\over 2}\pm\alpha}
e^{-i\pi(n+{1\over 2}\pm\alpha)\tau}
\nonumber \\
\chi_B^\dagger &=& \sum_{n\in Z} \chi^\dagger_{n+{1\over 2}\mp\alpha}
e^{-i\pi(n+{1\over 2}
\mp\alpha)\tau}\, .
\een
\nonumber \\
Note that in our notation
\be \label{en8}
(\chi_{n+{1\over 2}\pm\alpha})^\dagger =
\chi^\dagger_{-n-{1\over 2}\mp\alpha}\, .
\ee
The non-trivial anti-commutators are:
\be \label{en9}
\{ \chi_{n+{1\over 2}\pm\alpha}, \chi^\dagger_{-m-{1\over 2}\mp\alpha} \}
= \delta_{mn}\, .
\ee

In the Fock space built up from these oscillators, we need to project onto
states invariant under $(-1)^F$, $h$ and $\Omega$. From \refb{en1} we see
that both $(-1)^F$ and $h$ exchange the Chan Paton factors
$\sigma_3\mp i\sigma_2$. Since $\Omega$ acts on the Chan Paton factor
by transposition, it also exchanges the two sectors. Due to this reason
it is more convenient to take the independent
generators to be $(-1)^F$,  $(-1)^F h$ and $(-1)^F\Omega$. Since $(-1)^F h$
and $(-1)^F\Omega$ leave the Chan Paton factors untouched,
we can impose invariance under these transformations on the Fock space states,
and then take appropriate linear combinations of
the states from the two sectors
to get states invariant under $(-1)^F$. Let us for definiteness focus on the
sector with Chan Paton factor $\sigma_3 + i\sigma_2$. The
simplest way to implement
invariance under $(-1)^F h$ and $(-1)^F\Omega$ is to start from states at
$\alpha=0$ satisfying these projections and then simply follow them as we
continuously deform $\alpha$. During this process the oscillators
$\chi_{n+{1\over 2}}$ and $\chi^\dagger_{n+{1\over 2}}$ get deformed to
$\chi_{n+{1\over 2}+\alpha}$ and $\chi^\dagger_{n+{1\over 2}-\alpha}$
respectively, and the Fock vacuum $|0\rangle$ get deformed to $\vac_\alpha$
satisfying
\be \label{en10}
\chi_{n+{1\over 2}+\alpha}\vac_\alpha = 0, \qquad
\chi^\dagger_{n+{1\over 2}-\alpha}\vac_\alpha =0,
\qquad \hbox{for}\quad n\ge 0\, .
\ee
Notice that as $\alpha$ passes through (1/2), the state
$\vac_\alpha$ is no longer
the state with lowest $L_0$ eigenvalue. Instead for
$(3/2)>\alpha>(1/2)$ the state
with lowest $L_0$ eigenvalue, denoted by $\low$, is related to $\vac_\alpha$ by
the relation:
\be \label{en11}
\vac_\alpha = \chi^\dagger_{{1\over 2}-\alpha}\low\, ,
\ee
Note that $\low$ has opposite $(-1)^F h$ charge compared to $\vac$ since
$\chi$, $\chi^\dagger$ are odd under $(-1)^F h$.

Once we have determined the spectrum in each sector
invariant under $(-1)^Fh$ and
$(-1)^F\Omega$, we can find $(-1)^F$ invariant combination of states on the two
sectors by noting that under this transformation:
\ben \label{en12}
(\sigma_3 +i\sigma_2) &\to & (\sigma_3-i\sigma_2) \nonumber \\
\vac_\alpha &\to& -\vac_{-\alpha} \nonumber \\
\chi_{n+{1\over 2}+\alpha} &\to & \chi^\dagger_{n+{1\over 2}+\alpha}
\nonumber \\
\chi^\dagger_{n+{1\over 2}-\alpha} &\to & \chi_{n+{1\over 2}-\alpha}\, .
\een

We now note some important features of the spectrum.

\begin{itemize}

\item
At $\alpha=1$ the fermions $\chi$, $\chi^\dagger$ are again half
integer moded.  However, since the ground state $\low$ has opposite $(-1)^Fh$
charge compared to the vacuum $\vac$ at $\alpha=0$, this point is
not equivalent
to $\alpha=0$. In particular, if we denote by $X^\mu$, $\psi^\mu$
($1\le\mu\le 8)$
the bosonic
coordinates and their fermionic partners transverse to the
D-string, then the state
$\psi^\mu_{-{1\over 2}}\low$
is odd under $(-1)^F h$ and is projected out. On the other hand, if we
start from the state  $\psi^\mu_{-{1\over 2}}\vac$ at $\alpha=0$ and follow
it as $\alpha$ changes from 0 to 1, this state gets mapped to
$\psi^\mu_{-{1\over 2}} \chi^\dagger_{-{1\over 2}} \low$, and has dimension
1. Thus there are no
zero modes corresponding to separating the D-string and the anti- D-string
away from each other. This shows that at $\alpha=1$ (as well as at other
values of $\alpha$ which are not equivalent to $\alpha=0$) the D-string
anti- D-string system is bound.

$\chi$, $\chi^\dagger$ have half integer mode expansion again at $\alpha=2$.
Following an identical analysis one can easily verify that this time the new
ground state $\lowp$ has the same quantum numbers under various
projection operators
as in the case of $\alpha=0$ and hence this point is equivalent to $\alpha=0$.
In particular the set of states $\psi^\mu_{-{1\over 2}}\lowp$ survive all
projections and represent the freedom of
separating the D-string - anti-D-string
pair away from each other. This also shows that the D-string anti- D-string
system is most tightly bound ({\it i.e.} the mode corresponding to the
freedom of separating the pair has maximum mass) at $\alpha=1$. Thus we might
expect that the $\alpha=1$ point will represent the true minimum of the
tachyon potential when we increase the radius of $X$ away from $1/\sqrt 2$.
We shall see this more explicitly in section \ref{s4}.

\item
Note also that at $\alpha=1$, the ground state $\low$ is
invariant under $(-1)^Fh$.
This would correspond to a tachyonic mode. However, as we shall show now, this
state is projected out by the $(-1)^F\Omega$ projection. First note that since
this state can be written as $\chi_{1/2} \vac_{\alpha=1}$, it is the
image of the state $\chi_{-1/2}\vac$ at $\alpha=0$. The $(-1)^F$
invariant combination at $\alpha=0$, as seen from \refb{en12},
is given by
\ben \label{en12a}
&& {1\over\sqrt 2}
\{ \chi_{-{1\over 2}} \vac \otimes (\sigma_3+i\sigma_2)
- \chi^\dagger_{-{1\over 2}} \vac \otimes
(\sigma_3-i\sigma_2) \} \nonumber \\
&=& i (\psi_{-{1\over 2}} \vac \otimes \sigma_3 + \xi_{-{1\over 2}}\vac
\otimes \sigma_2)\, .
\een
The first term on the right hand side of this equation represents the
Wilson line of the U(1) gauge field on the D-string along the compact circle,
and is known to have the wrong $\Omega$ projection\cite{POLWIT}.
The second state
represents a state carrying momentum along $x$, but
also has wrong $\Omega$ projection since the Chan Paton factor is
anti-symmetric. Since by construction these states are even under $(-1)^F$,
we conclude that both are odd under $(-1)^F\Omega$, and hence are projected
out. Thus there is no tachyonic mode of the system at $\alpha=1$.

This analysis also shows us that if we consider an identical configuration
of D-string anti- D-string pair in type IIB string theory, then the
$\alpha=1$ state {\it does} have a tachyonic mode since there is no
$\Omega$ projection in this case. This tachyonic instability can be
traced to the fact that on (infinitely long) D-string
anti- D-string pair in type IIB string theory the tachyon
represents a complex field on the world volume,
and hence the minimum of the potential lies along a circle $T=T_0
e^{i\theta}$ instead of at two points $\pm T_0$. Since $\pi_0(S^1)$
is trivial, the kink solution is no longer topologically stable
(although it does represent a solution of the equations of motion)
and hence has a tachyonic mode representing decay into the
trivial solution $T=T_0$.

\item
Once we switch on the tachyon vev we break translational
invariance along $x$, and hence there must be a zero mode representing the
freedom of translating the soliton along $x$. The vertex operator for
this zero mode is constructed
as follows. Let $V_x$ denote the vertex
operator associated with this mode. If
$\delta_\epsilon$ denotes the transformation
$x\to x+\epsilon$, then, since the translation invariance is broken by the
tachyon vev, we would expect,
\be \label{en13}
\delta_\epsilon V_T \propto \epsilon V_x\, .
\ee
$V_T$ is the tachyon vertex operator defined in eqs.\refb{e18}-\refb{e20}.
This gives,
\be \label{en14}
V_x^{(-1)} = - e^{-\Phi_B} \xi_B \otimes \sigma_1, \qquad V_x^{(0)}
= \psi_B \eta_B \otimes \sigma_1\, .
\ee
This represents a marginal operator for all values of tachyon vev $\alpha$.

\item Note that the sectors with Chan Paton factors $I$ and $\sigma_1$
contain Fock space states carrying $((-1)^F, (-1)^Fh)$ values
(1,1) and $(-1,1)$ respectively, with the SL(2,R) invariant vacuum
assigned quantum numbers $(-1,-1)$. On the other hand, after switching
on the tachyon vev, the combined spectrum from the
sectors with Chan Paton factors
$\pmatrix{1 & \pm 1\cr \mp 1 & -1}$ contains Fock space states
carrying $((-1)^F, (-1)^Fh)$ values $(\pm 1, -1)$ {\it if we continue
to assign quantum numbers $(-1,-1)$ to the SL(2,R) invariant vacuum}.
Thus once we
combine the states from all four Chan Paton sectors, all Fock space
states are present in the spectrum before projection under $\Omega$.

\end{itemize}

Besides the modes discussed here, there are other
open string states whose one end lie on the (anti-) D-string and the
other end lie on the 9-brane. The effect of tachyon vev on the
spectrum of these states can be analysed by very similar method,
but we shall not do it here.

\sectiono{Deforming Away from the Critical Radius} \label{s4}

In this section we shall consider the effect of switching on the
perturbation that deforms the radius away from the critical radius.
In particular we shall show that:
\begin{itemize}
\item In the presence of this perturbation the tachyon develops a
tadpole except at $\alpha=0$ and at $\alpha=1$. Thus these are
the only two points which correspond to solutions of equations of
motion.

\item As we increase the radius of the circle in the $x$ direction,
the tachyon represents a state with positive mass$^2$
if $\alpha=1$. Thus $\alpha=1$ represents
a stable solution for $R>(1/\sqrt 2)$.

\item In the $R\to\infty$ limit we can represent the $\alpha=1$
solution as a D0-brane in type I string theory.

\end{itemize}

We begin by analysing the effect of radius deformation on the one point
function of the tachyon at an arbitrary value $\alpha$ of the
background tachyon field. If $V_r$ denotes the closed string
vertex operator associated with the radius deformation, then
the effect of switching on this perturbation on the tachyon
one point function is proportional to the two point function
$\langle V_r V_T\rangle_\alpha$ on the disk, with $V_T$ inserted at the
boundary of the disk and $V_r$ inserted in the interior of the disk.
For definiteness we shall take $V_T$
in the $-1$ picture and $V_r$ in
the $(-1,0)$ picture for this calculation.\footnote{$(-1,0)$ picture
means that we use $-1$ picture for the left-moving sector and 0
picture for the right-moving sector.} $V_T^{(-1)}$ is given in
eq.\refb{e21}, whereas using eqs.\refb{e15}-\refb{eyy3} we may express
$V_r^{(-1,0)}\propto e^{-\wt\Phi}\p X_R \wt\psi$ as
\be \label{en15}
V_r^{(-1,0)}
\propto e^{-\wt\Phi} \eta \Big( e^{i\sqrt 2 (\phi_L+\phi_R)}
+e^{i\sqrt 2 (\phi_L-\phi_R)} -e^{-i\sqrt 2 (\phi_L-\phi_R)}
-e^{-i\sqrt 2 (\phi_L+\phi_R)} \Big)\, ,
\ee
$\wt\Phi$ being the left-moving component of the
bosonized ghost\cite{FMS}.
The vertex operators $e^{\pm i\sqrt 2(\phi_L+\phi_R)}$ carry net
$\phi$ momentum. Since neither the tachyon vertex operator nor the
background tachyon field carry any $\phi$ momentum, and since the
Neumann boundary condition on $\phi$ conserves $\phi$ momentum, the
correlation function involving these two terms in $\langle V_r^{(-1,0)}
V_T^{(-1)}\rangle_\alpha$ vanish. Thus
the matter part of the correlation function is proportional to:
\be \label{en16}
\langle \eta(P)
\Big( e^{-i\sqrt 2 (\phi_L - \phi_R)} - e^{i\sqrt
2(\phi_L-\phi_R)}\Big)(P)
\eta_B(Q) Tr\Big(\sigma_1 e^{{i\over 2\sqrt 2}\alpha \sigma_1
\ointop dt \p_t \phi_B}\Big) \rangle\, ,
\ee
where $P$ denotes a point in the bulk, $Q$ denotes a point on the
boundary and $t$ parametrizes the boundary.
Note that we have included the effect of the tachyon background by
including the exponential factor inside the trace over the Chan Paton
factor, so $\langle ~\rangle$ in the above equation represents correlation
function in the theory without any tachyon vev.\footnote{Since
the tachyon vertex operator connects a D-string with $Z_2$ Wilson line
to an anti- D-string without a $Z_2$ Wilson line, there is some
subtlety in writing the contribution of the operators at the boundary as
a trace over Chan Paton factors. However, since the D-string and the
anti- D-string differ in their coupling to the
RR sector states, and also in their coupling to the states
carrying odd winding number along the $x$ direction
(see, for example \cite{BOUND}), this subtlety
does not affect the computation when the operator in the bulk is in
the NSNS sector and carries zero winding number along $x$.} The
trace over the Chan Paton factors can be easily evaluated and the
contribution may be written as
\be \label{en17}
i \langle \eta(P) \eta_B(Q)
\Big( e^{i\sqrt 2 (\phi_L - \phi_R)} - e^{-i\sqrt
2(\phi_L-\phi_R)}\Big)(P) \sin({1\over 2}\alpha\pi w_\phi)\rangle\, ,
\ee
where
\be \label{en18}
w_\phi = {\sqrt 2\over 2\pi} \ointop dt \p_t \phi_B\, ,
\ee
measures the total $\phi$ winding number carried by the operators in the
bulk. Since
$e^{\pm i\sqrt 2 (\phi_L - \phi_R)}$ carry $\phi$ winding number $\pm 2$
we see that this correlation function is proportional to
\be \label{en19}
i\sin(\alpha\pi) \langle \eta(P) \eta_B(Q)
\Big( e^{-i\sqrt 2 (\phi_L - \phi_R)} + e^{i\sqrt
2(\phi_L-\phi_R)}\Big)(P) \rangle\, .
\ee
The correlation function $\langle\eta(P) \eta_B(Q)\rangle$ is
non-vanishing on the disk.
On the other hand, due to Neumann boundary condition on $\phi$, the
expectation values of
$e^{\pm i\sqrt 2 (\phi_L - \phi_R)}(P)$ on the disk are non-vanishing
and equal. Thus we see that the one point function of the tachyon in
the presence of the radius perturbation is proportional to
$\sin(\alpha\pi)$. This vanishes only at $\alpha=0$ and $\alpha=1$
(other integer values of $\alpha$ being equivalent to
0 or 1). Thus only at these two values of
$\alpha$ we get solutions of the equations
of motion.

Next we analyse the two point function of the tachyon field at
$\alpha=1$. This
time we can take the two tachyon vertex operators to be in the $-1$ picture
and $V_r$ in the (0,0) picture, given by:
\ben \label{en20}
V_r^{(0,0)} &=& \p X_R \p X_L = -{1\over 2} \xi \eta \wt\xi \wt\eta
\nonumber \\
&=& {1\over 4} \eta \wt \eta
\Big( e^{-i\sqrt 2 (\phi_L - \phi_R)} + e^{i\sqrt
2(\phi_L-\phi_R)}
+ e^{-i\sqrt 2 (\phi_L + \phi_R)} + e^{i\sqrt
2(\phi_L+\phi_R)}\Big)(P)\, .
\een
Thus the matter part of the relevant correlation function is of the
form:
\ben \label{en21}
&& \langle \eta(P) \wt \eta(P)
\Big( e^{-i\sqrt 2 (\phi_L - \phi_R)} + e^{i\sqrt
2(\phi_L-\phi_R)}
+ e^{-i\sqrt 2 (\phi_L + \phi_R)} + e^{i\sqrt
2(\phi_L+\phi_R)}\Big)(P) \nonumber \\
&&  \qquad \eta_B(Q_1) \eta_B(Q_2)
Tr\Big(\sigma_1 \sigma_1 e^{{i\over 2\sqrt 2} \sigma_1
\ointop dt\p_t \phi_B}\Big) \rangle\, ,
\een
where again $P$ denotes a point in the interior of the disk and
$Q_1$ and $Q_2$ denote points on the boundary of the disk.
The trace over the Chan Paton factors gives
$\cos(\pi w_\phi/2)$. As in the
case of one point function, the contribution to the correlator
from the terms involving $\exp(\pm i\sqrt 2 (\phi_L+\phi_R))$ vanishes
due to $\phi$ momentum conservation.
For the other two terms carrying $w_\phi=\pm 2$, $\cos(\pi w_\phi/2)$
gives a contribution of $-1$. Thus the net effect of the tachyon vev
corresponding to $\alpha=1$ is to change the sign of this three point
function. Since we already know that in the absence of tachyon vev the
tachyon mass$^2$ {\it decreases} as we increase the radius, we conclude
that for $\alpha=1$ the tachyon mass$^2$ {\it increases} as we increase the
radius. Thus $\alpha=1$ represents a stable ground state of the system
for $R>(1/\sqrt 2)$. This analysis also shows that for
$R<(1/\sqrt 2)$ the tachyon mass$^2$ becomes negative at $\alpha=1$,
indicating that the spinor state becomes unstable in this range of
values of $R$.

We can generalise this analysis to include arbitrary number of
tachyon vertex operators on the boundary and arbitrary number of
insertions of $V_r^{(0,0)}$ in the interior. For $\alpha=1$, the
effect of the tachyon vev is to introduce a factor of
\be \label{enp1}
\exp({i\over 2} \pi w_\phi\sigma_1) = \cos(\pi w_\phi/2) + i \sigma_1
\sin(\pi w_\phi/2)\, ,
\ee
at the boundary. Since each term in $V_r^{(0,0)}$ carries even $w_\phi$
charge, the total $w_\phi$ carried by all the closed string vertex
operators is always even, and hence the contribution from the
$\sin(\pi w_\phi/2)$ term vanishes. The effect of the $\cos(\pi
w_\phi/2)$ term is to change the sign of every term in each
$V_r^{(0,0)}$ carrying odd $(w_\phi/2)$. Using expression \refb{en20}
for $V_r^{(0,0)}$ we see that it gets transformed to:
\ben \label{enp2}
&& {1\over 4} \eta \wt \eta
\Big( -e^{-i\sqrt 2 (\phi_L - \phi_R)} - e^{i\sqrt
2(\phi_L-\phi_R)}
+ e^{-i\sqrt 2 (\phi_L + \phi_R)} + e^{i\sqrt
2(\phi_L+\phi_R)}\Big)(P) \nonumber \\
&=& -{1\over 2} \eta \wt\eta \psi \wt\psi = -\p \phi'_R \p \phi'_L\, .
\een
Thus tachyon vev corresponding to $\alpha=1$
transforms the vertex operator $\p X_R\p X_L$ to $-\p \phi'_R \p \phi'_L$.
In other words, increasing the radius of the $x$ coordinate in the presence
of the tachyon vev corresponds to decreasing the radius of $\phi'$
coordinate with no tachyon vev.
If the $x$ radius is scaled up by a factor of
$L$, this would correspond to scaling down the $\phi'$ radius by a
factor of $L$. Thus an $x$ radius of $L/\sqrt 2$ will correspond to
a $\phi'$ radius of $1/(\sqrt 2 L)$.
Since the tachyon vertex operator
carries $\phi'$ momentum (as can be easily verified from eqs.\refb{e22a}
and \refb{e21}) a decrease in the $\phi'$ radius will increase the
mass$^2$ of the tachyon. This confirms our previous perturbative result.

The analysis can be easily extended to all NS sector vertex operators
with Chan Paton factors proportional to $\sigma_1$ or $I$. These
Chan Paton factors
commute with the Chan Paton factors associated
with the tachyon. Furthermore, since for correlation functions involving
NS sector states on the disk we have Neumann boundary condition on
$\phi$ ($\phi_L=\phi_R$) everywhere on the boundary, the vertex operators
inserted at the boundary may
carry $\phi$ momentum but no $\phi$ winding. Hence
insertion of these vertex operators do not create any discontinuity
in $\phi_B$ on
the boundary, and  $\ointop\p_t\phi_B$
along the boundary can be identified as $\sqrt 2\pi w_\phi$, with
$w_\phi$ measuring the total $\phi$ winding number of all the operators
inserted in the interior of the disk. Thus the effect of a
tachyon vev corresponding to $\alpha=1$ may again be represented by
\refb{enp1}. We can now repeat
the argument of the previous paragraph to show that the effect of
increasing the radius of the $x$ coordinate in the presence of the
tachyon vev is to effectively decrease the radius of the $\phi'$
coordinate.

Next we consider the case where on the boundary we have vertex
operators with Chan Paton factors proportional to $\sigma^\pm
=\sigma_3\pm i\sigma_2$. For illustration we consider the case
where we have two vertex operators at the boundary $-$ one proportional
to $\sigma^+$ and the other to $\sigma^-$ $-$ but
this construction can be easily generalized for any correlation
function with arbitrary insertions of NS sector open string vertex
operators at the boundary. For the two point correlator
the typical operator
insertions at the boundary take the form:
\be \label{enp4}
Tr\Big( \sigma^- e^{i k_1\phi_B(Q_1)} e^{{i\over 2\sqrt 2}
\sigma_1 \int_{Q_1}^{Q_2} \p_t\phi_B dt}
\sigma^+ e^{i k_2\phi_B(Q_2)} e^{{i\over 2\sqrt 2}
\sigma_1 \int_{Q_2}^{Q_1+2\pi} \p_t\phi_B dt} \Big)\, ,
\ee
where $t$ denotes the coordinate on the boundary of the disk with
periodicity $2\pi$, $Q_1$ and $Q_2$ represent the points on the boundary
where the vertex operators have been inserted, and $k_1$ and
$k_2$ are the
$\phi$ momenta carried by the vertex operators. Using the
commutation relations between the Pauli matrices, and the relations
\ben \label{enp5}
\int_{Q_1}^{Q_2} \p_t \phi_B dt &=& \phi_B(Q_2) - \phi_B(Q_1) \, ,
\nonumber \\
\int_{Q_2}^{Q_1+2\pi} \p_t \phi_B dt &=& \phi_B(Q_1) - \phi_B(Q_2)
+ \sqrt 2 \pi w_\phi \, ,
\een
one can bring \refb{enp4} to the form:
\be \label{enp6}
Tr\Big( \sigma^- e^{i (k_1 +{1\over \sqrt 2})
\phi_B(Q_1)}
\sigma^+ e^{i (k_2- {1\over \sqrt 2})
\phi_B(Q_2)} e^{i\pi w_\phi \sigma_1/2} \Big)\, .
\ee
Thus we see that the effect of switching on the tachyon vev is to shift
the $\phi$ momenta in vertex operators associated with $\sigma^\mp$
by $\pm 1/\sqrt 2$ $-$ a fact which has already been taken into account
in section \ref{s3} $-$ and simultaneously introduce a factor of
$\exp(i\pi w_\phi \sigma_1/2)$ at the boundary. The later factor
transforms the operators $\p X_R \p X_L$ inserted in the
interior of the disk to $-\p \phi'_R \p\phi'_L$ as before.
This shows that for all correlation
functions involving NS sector open string vertex operators, the effect
of deforming the $x$-radius after switching on the tachyon vev is to
change the $\phi'$ radius in the opposite direction.

{}From \refb{en14}, and the bosonisation rules \refb{eyy3} we see that
the vertex operator $V_x$ in the zero picture is proportional to
\be \label{enp3}
i \p \phi'_B \otimes \sigma_1\, ,
\ee
and hence represents a Wilson line along the $\phi'$ coordinate. This
clearly remains a marginal deformation even when we change the $\phi'$
radius $R_{\phi'}$.
This shows that when we change the radius of the $x$ coordinate
in the presence of tachyon vev, the translational zero mode
continues to represent a marginal deformation. Furthermore, this
zero mode, being a Wilson line along $\phi'$,
represents a periodic coordinate with periodicity
proportional to $1/R_{\phi'}$. Since $R_{\phi'}$ is proportional to
$(1/R)$, $R$ being the radius of the $x$ coordinate, we see that the
periodicity is proportional to $R$, as is expected of a zero mode
representing translation along a circle of radius $R$. In particular,
this shows that in the $R\to\infty$ limit the zero mode takes value
on a real line, representing the position of the soliton along the
$x$ axis.

In calculating the  spectrum in the $R\to\infty$ limit we can
simplify the analysis by making a T-duality
transformation in the coordinate $\phi'$. Since (after taking into
account the effect of the tachyon vev) the coordinate $\phi'$
has radius $1/(2R)$, the T-duality will give rise to dual
coordinate $\phi'_D$ of radius $2R$, and at the same time
transform the Neumann boundary condition on $\phi'$ to
Dirichlet boundary condition on $\phi'_D$. In the limit $R\to\infty$,
$\phi'_D$ has infinite radius, and we recover an extra
SO(9) symmetry that mixes the pair ($\phi'_D,\xi)$
with $(X^\mu,\psi^\mu)$ for $1\le\mu\le 8$.\footnote{Although
$(-1)^F$ and $h:X\to X +\sqrt 2 \pi$ projections act differently on
$(\phi'_D,\xi)$ and $(X^\mu,\psi^\mu)$, from the analysis of the
previous section we have seen that once we combine the states
from all four Chan Paton sectors, we have all the Fock space
states carrying $(-1)^F=\pm 1$ and $h=\pm 1$. Thus the full spectrum
is invariant under the exchange
$(\phi'_D,\xi)\leftrightarrow(X^\mu,\psi^\mu)$ before $\Omega$
projection. On the other hand, one can show by a detailed analysis
that the $\Omega$ projection also preserves the SO(9) symmetry of
the spectrum. The main point is to note that $\Omega$ induces a
transformation $\phi'\to -\phi'$, since otherwise the vertex
operator $V_x^{(0)}\propto \p_t \phi'_B\otimes \sigma_1$, involving
tangential derivative of $\phi'$ along the boundary, could not be
$\Omega$ invariant. This in turn implies that the dual coordinate
$\phi'_D$ does not change sign under $\Omega$, just as the $X^\mu$'s
for $1\le\mu\le 8$. Furthermore, since the vertex operators
$\xi_B\otimes \sigma_1$ and $\psi^\mu\otimes I$ are both present in the
spectrum in the $-1$ picture even after the $\Omega$ projection, the
oscillators of $\xi$ and $\psi^\mu$ transform under $\Omega$ in
identical manner.}
Thus we see that although
to start with we had broken the SO(9) symmetry of the problem by
identifying one special direction along which the D-string anti- D-string
pair lies, we recover the full SO(9) symmetry of the spectrum at
the end.

Next we turn to the fate of the Ramond sector states under radius
deformation. We shall focus on two point function involving two
Ramond sector vertex operators; this is sufficient for analysing
the spectrum. Since the insertion of a Ramond sector vertex
operators converts a Neumann boundary condition on $\phi$ to a
Dirichlet boundary condition, we can no longer represent the
tachyon vertex operator in the zero picture by \refb{e23} everywhere
in the boundary. However we can still express it as
\be \label{eps1}
V^{(0)}_T = i\sqrt 2 \p \phi_R \otimes \sigma_1\, .
\ee
Since the Ramond sector vertex operators do not carry either
momentum or winding along $\phi$, $\phi_R$ is continuous
across the Ramond sector vertex operators inserted at the
boundary. Following the same arguments as in the case of NS
sector states, we see that the effect of switching on the tachyon
vev on the correlation functions is to
\begin{itemize}
\item give a factor of
\be \label{eps2}
\exp\Big( {i\over\sqrt 2} \sigma_1 \ointop dt \p_t\phi_R\Big)\, ,
\ee
at the boundary, and
\item introduce factors of
\be \label{eps3}
\exp(\mp i\sqrt 2\phi_R)\, ,
\ee
at the location of Chan Paton factors $\sigma^\pm$. However, since
the Ramond sector vertex operators are
localized at $\phi=0$, \refb{eps3}
becomes trivial.
\end{itemize}
Let us now switch on the radius deformation. This corresponds to adding
a term proportional to integral of
$\p X_L \p X_R$ to the two dimensional action,
which in turn can be expressed as \refb{en20}. Using the operator
product expansion of scalar fields, it is easy to see that acting
on the right hand side of \refb{en20}, \refb{eps2} gives an
overall factor of $-1$. Thus for this correlation function, the
effect of the tachyon vev is to change the sign of the $\p X_L \p X_R$
term, {\it i.e.} to convert a perturbation corresponding to an increase
in radius to one corresponding to a decrease in the radius. In
particular in order to compute the spectrum of Ramond sector states
at $x$ radius $R=L/\sqrt 2$ in the presence of a tachyon vev $\alpha=1$,
we simply need to compute the spectrum at a new
$x$ radius $R_x=1/(\sqrt 2 L)$
in the absence of any tachyon vev. If we T-dualize in the $x$ coordinate,
and use the dual radius $x_D$ as the independent bosonic variable,
then the $x_D$ radius goes to $\infty$ as $R\to\infty$, and the Neumann
boundary condition on $x$ corresponds to Dirichlet boundary
condition on $x_D$. Thus we
recover the full SO(9) invariance of the spectrum in this
limit.\footnote{Again in this case we need to ensure that various
projections do not destroy the symmetry under the exchange of $x_D$
with one of the transverse coordinates $x^\mu$. To see this, let us
first note that in this case
the $h$ projection simply removes all states in Chan Paton sectors
$\sigma_1$ and $\sigma_2$, since states in this sector carry odd unit
of $x$ momentum, and hence become infinitely massive in the limit when
the effective $x$-radius goes to zero. Thus we are left with Chan
Paton factors $I$ and $\sigma_3$. The action of $\Omega$ on Ramond
sector states with Chan Paton factors $I\pm \sigma_3$, representing
wrapped D-branes and anti- D-branes respectively, can be shown to
differ by a factor of $-1$ using the results of \cite{GIMPOL}. Thus
when we combine the set of Fock space states from both sectors, we
can simply ignore the $\Omega$ projection. Finally, since $(-1)^F$
changes the sign of $\psi^\mu$ as well as of $\psi$, projection
under $(-1)^F$ does not destroy SO(9) invariance.}

We can now address the question of the dependence of the mass of the
soliton on the radius. In the absence of tachyon vev, the mass of the
D-string anti- D-string pair is expected to be proportional to $R$. Let
us first see how this arises in the string calculation. For this we
compute the interaction between a pair of such systems at large
separation by computing the partition function of open strings with
two ends lying on the two solitons, and equate it to
\be \label{enp4a}
G_N M^2 / l^6\, ,
\ee
where $G_N$ is the Newton's constant, $M$ is the mass of the
system, and $l$ is the separation. $G_N$ in the (8+1) dimensional theory
obtained by compactifying the $x$ direction is proportional to $(1/R)$.
On the other hand, for large separation $l$ between the two systems, the
leading contribution to the open string partition function comes from
the states with large $x$ momentum. The density of such states is
proportional to $R$. Thus we get,
$(M^2/R)\propto R$.
This gives $M\propto R$.

In the presence of tachyon vev, $G_N$ is still proportional to $(1/R)$.
The open string states in NS (R) sector are now better regarded as states
carrying definite momentum along $\phi'$ ($x$),
with $\phi'$ ($x$) having radius
proportional to $1/R$. For large $l$ the leading contribution
comes from states with large $\phi'$
($x$) momentum whose density of states
is proportional to $R_{\phi'}$ ($R_x$)
and hence to $R^{-1}$. Thus we get
$(M^2/R)\propto (1/R)$. This shows that $M^2$ is independent of $R$
as has been advertised earlier.

Let us now summarise the main result of this section. Starting
from the description of the SO(32) spinor of type I as a tachyonic
kink solution on the D-string anti- D-string pair, we have shown
that there is a description of these states in terms of an
exact boundary conformal field theory. In this description these
states are regarded as non-supersymmetric D0-branes of type I
string theory. There is no GSO projection on the open string
states\footnote{The possibility of existence of such D0-branes was
first mentioned by Bergman\cite{PRIV}.}. The NS sector ground state
of the open string represents a tachyonic mode, but this is
odd under the world sheet parity transformation $\Omega$ and is
projected out. Finally, the extra factor of $\sqrt 2$ in the
mass formula (eq.\refb{e27}) compared to the mass of an ordinary
D0-brane can be understood as follows. The open string partition
function of an ordinary D0-brane contains a projection operator
${1\over 2} (1+(-1)^F)$; of this the first term represents NSNS
exchange diagram, whereas the second term represents the RR
exchange diagram in the closed string channel. In the absence of
this projection operator the NSNS exchange diagram doubles, whereas
the RR exchange diagram is absent. This gives a net enhancement factor
of $\sqrt 2$ in the mass formula.\footnote{The mass is being measured
in the variable of the type IIB theory before the $\Omega$ projection;
hence in this calculation we do not take into account the extra factor
of $1/2$ coming from the $\Omega$ projection.}

\sectiono{Other Solitons on Brane-Antibrane Pair} \label{s5}

Encouraged by the fact that the tachyonic kink solution on the D-string
anti- D-string pair makes sense, we can try to look for other kinds
of solitons on the brane anti-brane pair. The first example that we
shall study is a vortex like solution on the membrane anti-membrane
pair in type IIA string theory. In this case the tachyon associated
with the open string stretched between the membrane and the
anti-membrane is a complex scalar field $T$, $T$ and $T^*$ being
associated with the two possible orientations of the open string.
Thus the manifold describing the minimum of the tachyon potential
has the structure of a circle, described by the equation:
\be \label{ela1}
|T| = T_0\, ,
\ee
for some real number $T_0$. The argument of \cite{TACH} shows that at
$|T|=T_0$ the negative contribution from the tachyon potential
exactly cancels the sum of the tensions of the membrane and the
anti-membrane, giving vanishing total energy density.

There is a $U(1)\times U(1)$ gauge field living on the world volume
of the membrane anti-membrane system. The tachyon carries one unit
of charge under each of these gauge fields. If $A_\mu^{(1)}$ and
$A_\mu^{(2)}$ denote the gauge fields coming from the D-brane and the
anti- D-brane respectively, then the kinetic term for the tachyon field
takes the form:
\be \label{ela2}
|D_\mu T|^2\, ,
\ee
where
\be \label{ela3}
D_\mu T = (\p_\mu - i A_\mu^{(1)} + i A_\mu^{(2)})T\, .
\ee
We can now consider a static,
finite energy vortex like configuration for the
tachyon field such that in the polar coordinates $(r,\theta)$ on the
membrane, the asymptotic field configuration takes the form:
\be \label{ela4}
T \simeq T_0 e^{i\theta}, \qquad A^{(1)}_\theta - A^{(2)}_\theta
\simeq 1\, , \qquad \hbox{as}\quad r\to\infty\, ,
\ee
so that both the kinetic and the potential terms vanish sufficiently
fast as $r\to\infty$. This soliton will describe a stable, finite
mass particle in type IIA string theory.

In order to identify this particle, let us note that
for this solution,
\be \label{ela4a}
\ointop (A^{(1)}-A^{(2)})\cdot dl = 2\pi\, ,
\ee
and hence it carries one unit of magnetic flux associated with the
gauge field $(A^{(1)}_\mu-A^{(2)}_\mu)$ on the world volume of the
membrane anti-membrane system. This in turn implies that it carries
one unit of $D0$ brane charge. Thus this soliton is indistinguishible
from the D0-brane of type IIA string theory, and is simply a different
representation of the same particle. Presumably this relation can be
proved explicitly by studying the boundary conformal field theory
describing this solution in a manner analogous to the previous two
sections.
The above construction can be trivially generalised to
represent the $p$-brane of type II string theory as a vortex solution on the
$(p+2)$-brane anti- $(p+2)$-brane pair.

The second example that we shall study is a string like solution
on the 5-brane anti- 5-brane system in type I string theory. The gauge
field living on the world volume of the system is
$SU(2)\times SU(2)$\cite{WITTSM,GIMPOL}, and the tachyon associated with
the open string stretched between the brane and the anti-brane is
a $2\times 2$ matrix valued field transforming in the $(2,2)$
representation of the $SU(2)\times SU(2)$ group. Thus if $U$ and $V$
denote the SU(2) matrices associated with the two gauge groups, then
\be \label{ela5}
T\to U T V^\dagger\, ,
\ee
under a gauge transformation.
The projection under the world-sheet parity transformation $\Omega$
reduces the number of real fields from 8 to 4, giving the following
restriction on the form of $T$:
\be \label{ela6}
T = \lambda W\, ,
\ee
where $\lambda$ is a real number and $W$ is a $2\times 2$ unitary
matrix. The manifold labelling the minimum of the tachyon potential
is described by the SU(2)$\times$SU(2) invariant equation,
\be \label{ela7}
\det(T) = \lambda_0^2\, ,
\ee
for some constant $\lambda_0$. This gives
\be \label{ela8}
\lambda=\lambda_0\, .
\ee
Thus this manifold is labelled by the SU(2) matrix $W$, and has the structure
of the SU(2) group manifold, namely $S^3$. By the result of ref.\cite{TACH},
on this manifold the total contribution to the energy density from the
tension of the brane and the anti-brane, and the tachyon potential
vanishes.

The kinetic energy of the tachyon field has the form:
\be \label{ela8a}
Tr ((D_\mu T)^\dagger (D^\mu T))\, ,
\ee
where
\be \label{ela9}
D_\mu T = \p_\mu T + i A_\mu^{(1)} T - i T A_\mu^{(2)}\, ,
\ee
$A_\mu^{(1)}$ and $A_\mu^{(2)}$ being the SU(2) gauge fields living
on the brane and the anti-brane.
Since a string like configuration in five dimensions has asymptotic
boundary $S^3$, we can get a finite energy, static,
string like solution in this (5+1)
dimensional field theory by imposing
the following asymptotic form for various fields:
\be \label{ela10}
T \simeq \lambda_0 U, \qquad A_\mu^{(2)}\simeq 0, \qquad A_\mu^{(1)}
\simeq i\p_\mu U U^{-1}\, ,
\ee
where $U$ is an SU(2) matrix valued function, corresponding to the
identity map (with unit winding number)
from the asymptotic boundary $S^3$ to
the SU(2) group manifold $S^3$. This makes both the kinetic and the
potential term vanish sufficiently rapidly at infinity so as to give a
finite energy configuration. Non-triviality of $\pi_3(S^3)$
guarantees topological stability of this solution.

In order to identify this solution, we note from the asymptotic form of
$A_\mu^{(1)}$ that it carries one unit of instanton number for the gauge
field $A_\mu^{(1)}$ living on the five brane. Since this is a source of
the D-string charge in type I string theory\cite{DOUGLAS} we see that
the soliton solution that we have constructed can be identified to a
D-string of type I string theory.

\noindent{\bf Acknowledgement:} I wish to thank O. Bergman, M.
Gaberdiel, S. Mukherji, K. Narain, N. Seiberg and E. Witten for useful
discussion. I would also like to thank the Institute for Advanced
Study, Princeton, International Center for Theoretical Physics, Trieste
and CERN, Geneva for hospitality during the course of this work.

\end{document}